\documentstyle[12pt]{l-aa}
\input{psfig}
\newfont{\mysybl}{Symbol at 12pt}

\newdimen{\plotbreite}
\newdimen{\plothoehe}

\begin{document}

\thesaurus{03(11.01.2 ; 13.07.2)}
\title{Constraints on the 3-30~MeV emission of Seyfert galaxies}
\author{M. Maisack        \inst{1}
\and    K. Mannheim       \inst{2}
\and    W. Collmar        \inst{3}}
\offprints{maisack@astro.uni-tuebingen.de}
\institute{Astronomisches Institut der Universit\"at T\"ubingen,
           72076 T\"ubingen, Germany
\and       Universit\"ats-Sternwarte, Geismarlandstr. 11, 37083 G\"ottingen, 
           Germany
\and       Max-Planck-Institut f\"ur extraterrestrische Physik,
           Postfach 1603, 85740 Garching, Germany}
\date{ ; }
\maketitle

\begin{abstract}

Seyfert galaxies have not been detected by COMPTEL on the Compton Gamma Ray 
Observatory (CGRO) in the energy 
range 0.75-3~MeV, placing upper limits on their emission which are more 
than an order of magnitude below previously reported detections. Here, we 
extend our previous work to the energy range 3-30~MeV. Again, we find no 
evidence for emission from a cumulative sample of X-ray bright Seyfert galaxies. 
We use the recent results on the extragalactic background at MeV energies to 
constrain the possible emission of these sources and their contribution to 
the cosmic extragalactic background (CXB). 
The lack of $\gamma$-rays from Seyfert galaxies
strongly argues against hadronic
cascades as the underlying radiation mechanism and, consequently,
against recent claims that Seyferts might produce a high-energy
neutrino background with an energy flux similar to that of the CXB.

\keywords{Galaxies: active ; Gamma rays: observations}
\end{abstract}

\section{Introduction}

The extragalactic sky at MeV energies, observed by CGRO COMPTEL and EGRET, 
has been found to be dominated by the 
radio loud class of blazars, not radio quiet objects such as Seyfert 
galaxies, as had been expected before the launch of CGRO (e.g. v. Montigny 
et al. 1995).
In a study of the X-ray brightest Seyferts at energies 0.75-3~MeV, Maisack 
et al. (1995, hereafter Paper~I) found no evidence for emission from 
Seyfert galaxies, deriving upper limits more than an order of magnitude 
below previously reported fluxes from balloon experiments (Perotti et al. 
1981a, 1981b). No Seyfert galaxy has been detected by EGRET, either (Lin et 
al. 1993). 

At the same time, OSSE observations ranging up to several 100~keV detected 
spectra significantly steeper than those found in the 
standard X-ray bands (Johnson et al. 
1994), suggesting a thermal nature of the high-energy emission. Zdziarski et al.
(1995, hereafter Z95) combined (non-simultaneous) Ginga and OSSE data of 9 Seyfert 
galaxies to derive a mean spectrum which is characterised by a continuum 
which falls off exponentially with an e-folding energy of several 100~keV, 
plus a Compton reflected component which contributes mainly between 10 and 
50~keV.  

Therefore, it is plausible that the 
primary hard X-rays are thermal X-rays from a hot corona
(Haardt \& Maraschi 1993) or other active regions above an accretion
disk (Stern et al. 1995) which acts as the cold reflector. 
A Compton reflected 
component has been observed in many Seyfert galaxies 
(e.g., Nandra and Pounds 1994).  However,
a nonthermal origin of the {\it primary} emission is still not ruled out, 
since the
nonthermal spectrum could turn over at $\sim$100~keV as well or, if it
extends to $\gamma$-ray energies, it could be anisotropic. 
Therefore, this radiation could go unobserved in most sources, and be 
mainly visible
as reprocessed radiation.  
A likely candidate for such anisotropic emission 
are primary X-rays from a misdirected X-ray jet. 
Klein-Nishina
driven nonthermal emission
could also be highly anisotropic irradiating the
disk  (Ghisellini et al. 1991).  Analogously, photo-pair or photo-pion
driven hadronic cascades arising in a turbulent, low Mach number jet
also produce much more flux in the direction of the disk than
toward the observer (Mannheim 1995a).  The cascade
spectrum reaches up to a few MeV,
and a diffuse high-energy neutrino background with an energy flux
comparable to the CXB is expected from such
hadronic models (Stecker et al. 1991).
We probe the presence of a radiation component peaking
at high energies by direct COMPTEL measurements
employing the method of Paper~I to the 3-30~MeV high energy range.

The remainder of the paper is organised as follows: in sections 2 and 3, we 
describe the instrument and data analysis, and give the results. We then 
discuss the upper limits in connection to the recent COMPTEL results of the 
CXB.

\begin{table*}[th]
\renewcommand{\arraystretch}{1.37}
\begin{tabular}[c]{lrrlllll}
\multicolumn{8}{l}{Table 1. Individual 2$\sigma$ upper limits for Seyfert galaxies 
observed by COMPTEL in Phase I} \\
\hline
Source Name & l & b & Viewing Periods & UL 0.75-1~MeV$^a$ & 
 UL 1-3~MeV$^a$ & UL 3-10~MeV$^a$ & UL 10-30~MeV$^a$\\
\hline
NGC 7314      &  26.44  & -59.17 & 42             & 108.8 & 11.7 & 0.85 & 0.14\\
NGC 6814      &  29.15  & -16.01 & 7.5, 13, 43    &  49.0 &  4.7 & 0.79 & 0.08\\
NGC 5548      &  31.94  &  70.49 & 24             & 140.0 & 15.0 & 0.99 & 0.11\\
MRK 509       &  35.97  & -29.85 & 7.5, 13, 19, 43&  40.4 &  5.5 & 0.47 & 0.07\\
3C 445        &  61.87  & -46.71 & 19             &  50.0 &  5.0 & 1.09 & 0.07\\
MCG -2-58-22  &  64.08  & -58.76 & 19             &  37.5 &  7.5 & 1.13 & 0.08\\
NGC 7469      &  83.09  & -45.46 & 19, 28, 37     &  32.4 &  7.2 & 1.05 & 0.10\\
III ZW 2      & 106.97  & -50.63 & 28, 37         &  57.6 &  8.0 & 0.32 & 0.05\\
MRK 335       & 108.76  & -41.42 & 28, 37         & 106.8 & 12.4 & 0.55 & 0.03\\
NGC 4051      & 150.12  &  69.71 & 4              &  39.6 &  4.8 & 0.59 & 0.06\\
NGC 4151      & 155.05  &  75.06 & 4              &  15.8 &  4.8 & 1.10 & 0.13\\
3C 111        & 161.67  &  -8.81 & 15, 31, 36, 39 &  22.8 &  3.0 & 0.31 & 0.07\\
MCG +8-11-11  & 165.72  &  10.40 & 30, 36, 39     &  58.2 &  8.8 & 0.38 & 0.05\\  
NGC 1068      & 172.10  & -51.93 & 21             &  51.0 &  7.0 & 0.69 & 0.08\\
3C 120        & 190.37  & -27.39 & 1, 2.5         &  27.3 &  3.8 & 0.83 & 0.15\\
MCG +5-23-16  & 200.82  &  46.46 & 40             &  42.1 &  5.5 & 0.81 & 0.06\\
ARK 120       & 201.69  & -21.12 & 1, 2.5         &  50.9 &  6.6 & 1.11 & 0.11\\
NGC 3227      & 217.01  &  55.44 & 40             &  40.6 &  9.4 & 1.27 & 0.07\\
NGC 3783      & 287.45  &  22.94 & 12, 14, 32     &  22.2 &  3.3 & 0.49 & 0.07\\
FAIRALL 9     & 295.07  & -57.82 & 10             &  45.2 &  5.9 & 0.59 & 0.06\\
NGC 4593      & 297.47  &  57.40 & 3, 11          &  71.4 &  3.3 & 0.89 & 0.07\\
IC 4329 A     & 317.49  &  30.92 & 12             &  50.5 &  6.1 & 0.64 & 0.08\\
ESO 141-55    & 338.18  & -26.71 & 35, 38         &  83.2 & 16.0 & 1.49 & 0.11\\
MCG -6-30-15  & 338.46  &  44.53 & 12             &  85.5 & 10.6 & 0.74 & 0.09\\
NGC 5506      & 339.14  &  53.80 & 3, 11, 24, 25  &  30.5 &  5.2 & 0.43 & 0.07\\
NGC 7582      & 348.07  & -65.69 & 9, 42          & 122.2 &  9.9 & 1.71 & 0.12\\
\hline
\multicolumn{6}{l}
{$^a$ in 10$^{-8}$ photons cm$^{-2}$ s$^{-1}$ keV$^{-1}$} \\
\end{tabular}
\end{table*}

\section{Instrument and Data Analysis Methods}

The Compton telescope {\it COMPTEL} (Sch\"onfelder et al. 1993), as 
part of CGRO, 
covers the energy range 0.75-30 MeV. An all-sky survey in this energy range 
was conducted during the first 15 months of the GRO mission. We have overlaid 
50 observations of 26 individual X-ray prominent Seyfert galaxies obtained 
during this survey to improve the sensitivity over that of a standard two 
week observation. Relevant details about the instrument, 
data acquisition and the data analysis 
methods applied for this work have been described in detail in Paper~I.

\section{Observations and Results}
We have used the same set of observations as in the analysis of the 
0.75-3~MeV range reported in Paper~I to be able to directly 
compare the results at 3-30~MeV to those in the lower energy bands. 

Again, we detect no individual Seyfert galaxy, and no significant flux in 
the cumulative data which correspond to a net observation 
time of $\approx$ 70~days. The 2$\sigma$ 
upper limits for the individual sources are 
listed in Table 1, together with those for the two low energy bands already 
presented in Paper~I. 
The limits on the complete sample are 1.05 $\times$ 
10$^{-9}$ photons  cm$^{-2}$ 
s$^{-1}$ keV$^{-1}$ for the 3-10 MeV band 
and 1.32 $\times$ 10$^{-10}$ photons 
cm$^{-2}$ s$^{-1}$ keV$^{-1}$ in the 10-30 MeV band, and about the same in 
$\nu F_{\nu}$ as the limit in the 1-3~MeV band 
(9.7 $\times$ 10$^{-9}$ photons cm$^{-2}$ s$^{-1}$ keV$^{-1}$), see Paper~I). 
The 2$\sigma$ upper limits are shown 
in Fig.1, together with the extension of the average spectrum of Seyfert 
galaxies at several 100~keV derived from OSSE observations (Johnson et al. 
1994, Z95).

\begin{figure*}[tb]
  \psfig{figure=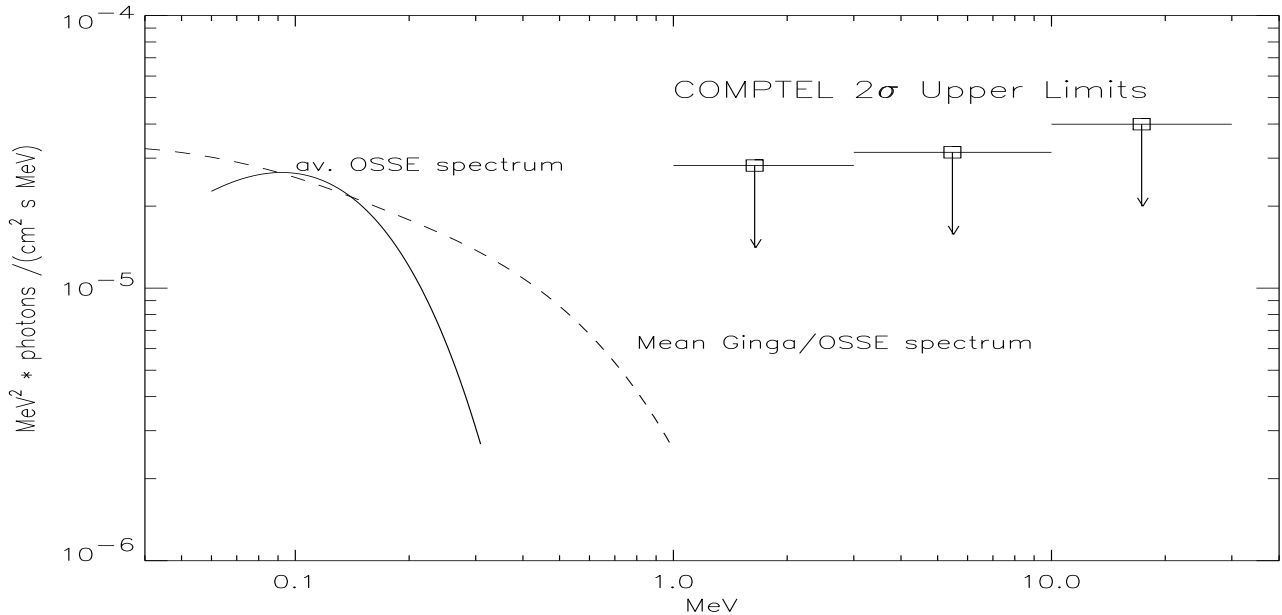,width=\textwidth,height=8.7cm}
  \caption[]{COMPTEL 2$\sigma$ Upper Limits Compared to X-Ray Samples of 
Johnson et al. (1994) and Z95}
\end{figure*}

This result is not surprising when compared to the analysis of Z95
who analysed (non-simultaneous) Ginga and OSSE data of 9 Seyfert galaxies  
and found that the average spectra can be 
described by an exponentially truncated power law with photon indices 
around the canonical X-ray values of 1.8-2.0 and an e-folding energies of 
several 100~keV. The OSSE data alone indicate even steeper spectra with 
e-folding energies of $\approx$~40-50~keV (Johnson et al. 1994). Simultaneous 
broad-band observations with XTE and OSSE could better constrain this 
parameter in the future. The best fit 
spectrum for the complete sample of Seyferts from Z95 
is shown in Fig.1 for comparison. It is obvious that the upper limits we 
derive are still substantially above the level of emission expected from 
the work of Z95.

It has been suggested that the thermal appearance of the Seyfert X-ray 
spectra could be due to a highly 
anisotropic nonthermal source emitting most photons toward the disk,
so that the observed spectrum is mostly due to reflection and 
Comptonization by the disk (Mannheim 1995a).   
A natural anisotropy of this kind develops for pair cascades
produced by ultrarelativistic protons (Lorentz factor $\gamma_{\rm p}$)
accelerated in a magnetized disk wind
('hadronic jet') as they cool by photo-production of secondary 
particles in the  
radiation field of the disk.   Pions are produced by head-on
collisions with UV photons (energy $\epsilon$) from the inner disk.  
In the rest frame
of the proton, the UV photons appear with energies $\epsilon'=
2\gamma_{\rm p}
\epsilon$ which leads to
catastrophic energy losses when $\epsilon'\ga m_\pi c^2$,
thereby stopping further
proton acceleration.  The pions subsequently decay giving rise to
an anisotropic cascade irradiating the disk hemisphere.   
Infrared photons
originating in the heated dust torus surrounding the central object
appear with energies $\epsilon'\ga 2m_{\rm e}c^2$ in the proton
rest frame giving rise to Bethe-Heitler $e^\pm$ pairs.  Owing to
the solid angle $\sim 2\pi$ subtended by the infrared photons,
the Bethe-Heitler pair distribution is nearly isotropic.
The pairs produce synchrotron $\gamma$-rays in the 3-10~MeV energy
range.
The expected energy flux in the $3-10$~MeV range
is maximally of the same order as the Compton reflected component,
but does not contradict the already derived flux limits at $\sim$MeV
due to the rather flat spectrum (Fig.2).  From our
observation alone, no constraints on this extra component
can be derived, as the upper limits are of about the same magnitude 
in $\nu F_{\nu}$ as the fluxes of the OSSE observations at several tens of 
keV (Z95).

Tighter constraints on the MeV emission of 
Seyferts than those obtained from the individual and cumulative 
observations can be 
derived from the recent results on the extragalactic background derived 
by Kappadath et al. (1996), who found that the MeV bump in the background 
(e.g. Gruber 1992) was an artifact owing to the detector background caused 
by charged particles being dependent on 
the geomagnetic rigidity at which individual measurements were conducted. 
Kappadath et al. (1996) conclude that there is no MeV bump, and that the 
spectrum of the CXB can be described by a power law from 100~keV to 
hundreds of MeV, as originally found by Mazets et al. (1975). The photon 
index of the power law connecting hard X-ray and MeV $\gamma$-rays 
lies in the range 2.5-3, significantly flattening toward EGRET
energies where $s=2.07\pm 0.03$ (Kniffen et al. 1996).

Recent modeling of the CXB (Comastri et al. 1995, Z95) 
has shown that the CXB at several tens of keV can be described by the 
superposition of Seyfert galaxies (more generally radio quiet AGN) 
at various levels of obscuration. 
Assuming similar spectra for all Seyferts from hard X-rays to MeV energies, 
the steep spectrum of the CXB places stronger constraints on the 
$\gamma$-ray flux from Seyferts than the actual observations reported in this 
paper. 

This can be seen from Fig.2, which shows the 2$\sigma$ 
upper limits derived from the 
COMPTEL data together with the average X-ray spectrum of Seyferts from Z95, 
plus the spectrum of the CXB (following Mazets 1975, Kappadath 1996 and 
Kniffen 1996) scaled to the Z95 spectrum. Even neglecting the possible 
contribution of other AGN and Supernovae type Ia to the CXB, the persistent 
emission from Seyferts must lie a factor of $\approx$ 10 below the COMPTEL 
upper limits to be consistent with the CXB. Accordingly, these constraints 
imply that the anisotropic nonthermal cascades
either have extremely large
disk/observer flux ratios obtaining values $\ga 10$ or, more
likely, that they
do not contribute substantially 
to the reflected component. 
This is in agreement with the non-detection of 
$>$~TeV neutrinos with the Fr\'ejus proton-decay experiment (Mannheim
1995b). 
A diffuse
neutrino background with an energy flux comparable to the CXB as proposed
by Stecker et al. (1991) therefore cannot be expected from radio-quiet AGN.

The main contribution 
to the primary X-ray emission from radio-quiet AGN
responsible for the reflection hump
seems to come from  coronal plasma near the inner
accretion disk (e.g., Haardt and Maraschi 1993)
or from a nonthermal (non-cascade) source with an intrinsic
turnover at $\sim 100$~keV.

\section{Future observations}

The next generation $\gamma$-ray mission INTEGRAL (e.g. Winkler et al. 1994) 
is supposed to provide a sensitivity at several 100~keV to 
1~MeV which will be a factor of 10 
below that of COMPTEL. Given the tight limits which the CXB measurements 
put on the average emission of Seyferts, this may be marginally sufficient 
to probe deep enough to detect Seyfert galaxies at this energy. 

\section{Summary}

Extending our work on Seyferts in the 0.75-3~MeV range, we find that 
Seyferts are not bright in the 3-30~MeV range, either.  An emission component
in this high energy range could have been expected from the $ \sim 5$~MeV bump
in the diffuse $\gamma$-ray
background as it was obtained from Apollo data by Trombka et al. (1977)
and from nonthermal models.
The null-result is in accord with the absence of a
bump in the diffuse background spectrum as it has recently
been obtained from COMPTEL data
by Kappadath et al. (1996).  The diffuse $\gamma$-ray background implies
that the average emission from Seyfert galaxies still lies a factor
of $\sim 10$ below the current upper limits for selected X-ray bright Seyferts.
In particular, since we do not find
evidence for nonthermal X-ray and $\gamma$-ray emission from Seyfert galaxies,
predictions of high-energy neutrinos from Seyferts 
(Stecker et al. 1991) are invalidated.

\begin{figure}[htb]
  \psfig{figure=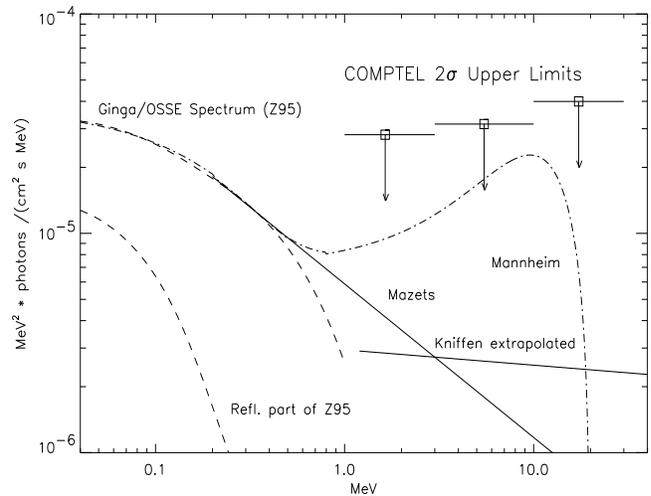,width=0.5\textwidth,height=7cm}
  \caption[]{Comparison of COMPTEL 2$\sigma$ Upper Limits to those derived from the 
CXB (solid lines) which is scaled to the keV emission (dashed lines) 
derived by Z95. The dashed-dotted line represents the maximal flux expected 
from a Bethe-Heitler pair component in the hadronic jet model of 
Mannheim (1995a)}
\end{figure}

\acknowledgements

MM acknowledges support by DARA grant 50~OR~92054. The 
COMPTEL project is 
supported by the German government through DARA grant 50 QV 90968.

\end{document}